# DATA MINING ATTRIBUTE SELECTION APPROACH FOR DROUGHT MODELLING: A CASE STUDY FOR GREATER HORN OF AFRICA

Getachew B. Demisse, Tsegaye Tadesse and Yared Bayissa

National Drought Mitigation Center, School of Natural Resources, University of Nebraska-Lincoln, Hardin Hall, 3310 Holdrege Street, P.O. Box 830988, Lincoln, Nebraska 68583-0988, U.S.A.

*ABSTRACT*

*The objectives of this paper were to 1) develop an empirical method for selecting relevant attributes for modelling drought and 2) select the most relevant attribute for drought modelling and predictions in the Greater Horn of Africa (GHA). Twenty four attributes from different domain areas were used for this experimental analysis. Two attribute selection algorithms were used for the current study: Principal Component Analysis (PCA) and correlation-based attribute selection (CAS). Using the PCA and CAS algorithms, the 24 attributes were ranked by their merit value. Accordingly, 15 attributes were selected for modelling drought in GHA. The average merit values for the selected attributes ranged from 0.5 to 0.9. The methodology developed here helps to avoid the uncertainty of domain experts' attribute selection challenges, which are unsystematic and dominated by somewhat arbitrary trial. Future research may evaluate the developed methodology using relevant classification techniques and quantify the actual information gain from the developed approach.*

*KEYWORDS*

*Attribute, CAS, Data mining, Modelling, PCA*

## 1. INTRODUCTION

Attribute selection is the process of identifying relevant information and removing as much of the irrelevant and redundant information as possible [1]. Attribute selection is also defined as "the process of finding a best subset of features, from the original set of features in a given data set, optimal according to the defined goal and criterion of feature selection (a feature goodness criterion)" [2]. In this paper, attribute selection is the process of selecting relevant drought variables for constructing drought prediction models in space-time dimensions. The attribute selection approach here is focused on identifying and selecting the most relevant drought descriptor variables from different sources and removing the irrelevant attributes without loss of information. In the context of the current research, redundant or irrelevant features are two distinct notions, since one relevant feature may be redundant in the presence of another relevant feature with which it is strongly correlated [3].

Attribute selection techniques can be categorized according to a number of criteria. One popular categorization has coined the terms *filter* and *wrapper* to describe the nature of the metric used to evaluate the worth of attributes [4]. Wrappers evaluate attributes by using accuracy estimates provided by the actual target learning algorithm. Filters, on the other hand, use general characteristics of the data to evaluate attributes and operate independently of any learning algorithm [1].

DOI: 10.5121/ijdkp.2017.7401          1

The assumption in machine learning and data mining models is that the attributes used for training are relevant to the target attribute being explained and the learning algorithms are designed to learn with the most appropriate attributes to use for making their decisions [5]. In addition to this, there should not be duplication of information in the attributes being used for the modelling experiment [1, 6].

In the past, the issue of attribute selection for developing data mining models was found to be unsystematic and dominated by arbitrary trial [7]. This is the major cause of model uncertainties [8, 9] and challenges for converting the theoretical models into practical real world problem solving applications [8]. It is also important to note that one of the most important tasks in data mining experiments is attribute selection [10]. This is because the output of a data mining experiment is highly dependent on the input attributes and their data values [8, 9]. For instance, one of the challenges in nearest neighbour data mining models is that the models are adversely affected by the presence of irrelevant attributes. All attributes are taken into account when evaluating the similarity of two cases, and irrelevant attributes introduce a random factor into this measurement. As a result, composite models are most effective when the numbers of attributes are relatively small and all attributes are relevant to the prediction task [11].

Witten et al. [5] indicated that decision tree methods choose the most promising attribute to split on at each point and should in theory never select irrelevant or unhelpful attributes. Contrary to this, adding irrelevant or distracting attributes to a dataset often confuses machine learning systems. Experiments with a decision tree learner (C4.5) have shown that adding to standard datasets a random binary attribute generated by tossing an unbiased coin impacts classification performance, causing it to deteriorate (typically by 5-10% in the situations tested). Specifically, instance-based learners are very susceptible to irrelevant attributes because they always work in local neighbourhoods, taking just a few training instances into account for each decision [5].

Practical machine learning algorithms (including top-down induction of decision tree algorithms such as CART and instance-based algorithms) such as instance-based learner (IBL) [12] are known to degrade in performance (prediction accuracy) when faced with many features that are not necessary for predicting the desired output [12]. On the other hand, algorithms such as Naive Bayes [12] are robust with respect to irrelevant features (i.e., their performance degrades very slowly as more irrelevant features are added), but their performance may degrade quickly if correlated attributes are added (if there is duplication of information), even if the attributes are relevant [5, 12]. In the past, in addition to relevancy of the attributes for modelling experiments, the redundancy of the attributes due to their high intercorrelation is not checked [2, 8, 9]. Because of the redundancy of the attributes, most of the classification algorithms suffer from extensive computation time, decrease in accuracy, and uncertainty in the interpretations of the model outputs [2].

It was also confirmed that in a given database, if too much irrelevant and redundant information is present then learning during the training phase is more difficult, and at the same time the redundant data directly lead to the problem of overfitting; as a consequence, the overall performance of the system will degrade [13].

The goal of this study was to develop an attribute selection method with special emphasis on drought modelling and prediction. The specific objectives of the paper were to: 1) develop an empirical method for selecting relevant attributes for modelling drought and 2) select the most relevant attribute for drought modelling and predictions in the GHA. In this paper, we focus on attribute selection techniques that produce ranked lists of attributes. These methods are useful for improving the performance of learning algorithms, and the rankings they produce can also provide the data miner with insight into their data by clearly demonstrating the relative merit of individual attributes [1]. Materials and methods for attribute selection approaches (principal components analysis and correlation-based attribute selection) are presented in detail in section 2. Section 3 discusses the major findings, and section 4 presents the conclusions.

## 2. MATERIALS AND METHODS

### 2.1 DATA SOURCE

The experimental dataset for this research was extracted from the Greater Horn of Africa (GHA), which is geographically located between -12.34°S to 35.7°N latitude and 21.1° to 51.5°E longitude. Administratively, the study area includes Burundi, Djibouti, Eritrea, Ethiopia, Kenya, Rwanda, Somalia, South Sudan, Sudan, Tanzania, and Uganda [14] (Figure 1). The region is known to have highly diversified topography and ecosystems [15]. The data extracted from this region is assumed to represent the diversified topography and ecosystems.

A total 24 different attributes from different domain areas were used for this experimental analysis. Most of the attributes were obtained from the National Oceanic and Atmospheric Administration (NOAA) data portal (https://www.esrl.noaa.gov/psd/data/climateindices/list/) [16]. The remaining attributes were obtained from the United States Geological Survey (USGS) [17], European Space Agency (ESA) [18], International Soil Reference and Information Centre (ISRIC)–World Soil Information [19], Climate Hazards Group Infrared Precipitation with Stations (CHIRPS) [20], EROS Moderate Resolution Imaging Spectroradiometer (eMODIS) [21], National Aeronautics and Space Administration (NASA) EARTHDAT [22], and The Nature Conservancy's (TNC) GIS data portal [23]. Table 1 presents the attributes, data types, and sources of the data for each attribute. The domain explanations, detailed attribute descriptions, and data preparation is available in the references listed. From each of the attributes listed in Table 1, time series data from 2001 to 2015 were extracted and used in the experimental analysis.

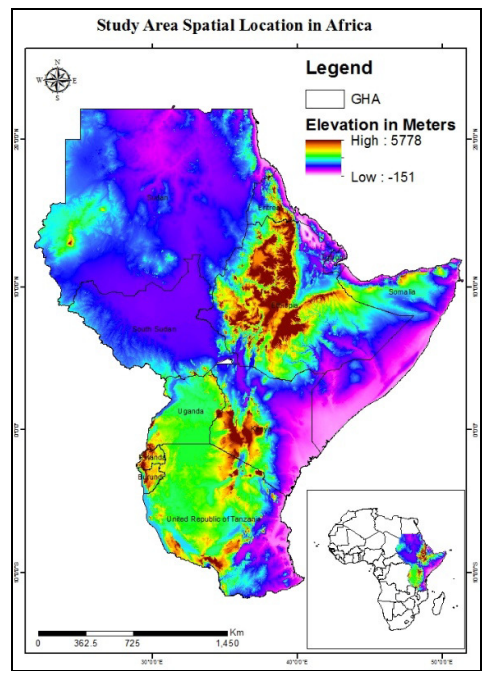

Figure 1: Location of the study area in Africa. The background map in this figure is elevation in meters.

International Journal of Data Mining & Knowledge Management Process (IJDKP) Vol.7, No.4, July 2017

Table 1. Attributes used for the experimental analysis.

| Attribute | Acronym | Type | Source and references |
|---|---|---|---|
| Digital Elevation Model | DEM | Biophysical | USGS [17] |
| Land Cover | LC | Biophysical | ESA [18] |
| Water Holding Capacity of the Soil | WHC | Biophysical | ISRIC - World Soil Information [19] |
| Normalized Precipitation | N_Precip | Climate | Climate Hazards Group Infrared Precipitation with Stations (CHIRPS) [20] |
| Standardized Seasonal Greenness | SSG | Satellite | eMODIS [21] |
| Soil Moisture | SM | Biophysical | NASA EARTHDAT [22] |
| Ecoregion | EC | Biophysical | The Nature Conservancy's (TNC) GIS data portal [23] |
| Atlantic Meridional Mode | AMM | Oceanic / Atmospheric | NOAA [16] |
| Dipole Mode Index | DMI | Oceanic / Atmospheric | NOAA/ESRL [24] |
| Multivariate ENSO Index | MEI | Oceanic / Atmospheric | NOAA [16] |
| Oceanic Niño Index | ONI | Oceanic / Atmospheric | NOAA [16] |
| Pacific Decadal Oscillation | PDO | Oceanic / Atmospheric | NOAA [16] |
| Trans-Niño Index | TNI | Oceanic / Atmospheric | NOAA [16] |
| Tropical Northern Atlantic Index | TNA | Oceanic / Atmospheric | NOAA [16] |
| Tropical Southern Atlantic Index | TSA | Oceanic / Atmospheric | NOAA [16] |
| Quasi-Biennial Oscillation | QBO | Oceanic / Atmospheric | NOAA [16] |
| Solar Flux (10.7cm) | SFLUX | Oceanic / Atmospheric | NOAA [16] |
| Central Tropical Pacific SST | Nino 4 | Oceanic / Atmospheric | NOAA [16] |
| East Central Tropical Pacific SST | Nino 3.4 | Oceanic / Atmospheric | NOAA [16] |
| Bivariate ENSO Timeseries | BEST | Oceanic / Atmospheric | NOAA [16] |
| Atlantic Multidecadal Oscillation | AMO | Oceanic / Atmospheric | NOAA [16] |
| North Atlantic Oscillation | NAO | Oceanic / Atmospheric | NOAA [16] |
| Pacific North American Index | PNA | Oceanic / Atmospheric | NOAA [16] |
| Southern Oscillation Index | SOI | Oceanic / Atmospheric | NOAA [16] |

## 2.2 ATTRIBUTE SELECTION APPROACHES

Our experiment for the best attributes selection involved searching through all possible combinations of attributes in the data to find which subset of attributes works best for the drought prediction. This process was done in two steps: 1) set up an attribute evaluator, and 2) determine the relevant search method. Setting up the evaluator determines the type of method to be used to assign a worth to each subset of attributes, and the search method determines what style of search is performed [25].

For the experimental analysis, we used Weka2 (Waikato Environment for Knowledge Analysis) [26], a powerful open-source Java-based machine learning workbench that can be run on any computer that has a Java runtime environment installed [1]. Weka has several machine learning algorithms and tools under a common framework with an intuitive graphical user interface. For this study, two modules in Weka were used: a data exploration module and attribute selection module [26].

Four attribute selection algorithms were considered for the current study: CAS, PCA, ReliefF feature selection algorithm, and wrapper (WrapperSubsetEval) [5]. These four algorithms were found to have better performances than other classic algorithms, such as the information gain algorithm [1]. In addition, these four algorithms were found to be able to handle a continuous dataset, which helped us not to discretize the data for our experiment analysis.

During the experimental analysis, the ReliefF feature selection algorithm and WrapperSubsetEval algorithms were found to be slow to handle the 24 attributes with huge datasets (about 519,000 records). Other research [1, 6] also noted that these two algorithms are too slow to execute on huge datasets and are not recommended for huge dataset manipulations. Specifically, Hall and Holmes [1] indicated that if the training dataset is large, the application of the wrapper approach may be unrealistic because of the enormous computation time required. Therefore, due to the

International Journal of Data Mining & Knowledge Management Process (IJDKP) Vol.7, No.4, July 2017

intensive computations and very long time needed to get the outputs of the analysis, the ReliefF and WrapperSubsetEval algorithms were not used in the current research. The details for the PCA and CAS attributes selection approaches are presented in the following subsections.

### 2.2.1 Principal Component Analysis (PCA)

The PCA is a technique that uses a linear transformation to form a simplified dataset retaining the characteristics of the original dataset [27-29]. If the original data matrix contains d dimensions and n observations, it is required to reduce the dimensionality into a k dimensional subspace. This transformation (equation 1) [27] is given by:

$$Y = E^T_{dxk} X_{dxn} \tag{1}$$

where $Y$ is a matrix of principal components, $E^T_{d \times k}$ is the matrix of standardized observational data (the projection matrix that contains $k$ eigenvectors corresponds to the $k$ highest eigenvalues), and $X_{d \times n}$ is matrix of eigenvectors (a mean-centered data matrix). Detailed descriptions of the theoretical and applications of PCA are presented in [27].

The goal of PCA is to extract the important information from the data table and express this information as a set of new orthogonal variables called principal components [27-29]. PCA also represents the pattern of similarity of the observations and variables by displaying them as points in maps [27]. The PCA algorithm was found to be useful in looking at the subset of attributes for capturing the variance in the identified principal axis, which helped us to see the associations of the attributes in explaining the target attribute (the SSG in the current experimental analysis).

### 2.2.2. Correlation-based Attribute Selection (CAS)

CAS is one of the most popular attribute subset selection methods [1, 6, 30]. The main objective of CAS is to obtain a highly relevant subset of features that are uncorrelated to each other [6]. In this way, the dimensionality of datasets can be drastically reduced and the performance of learning algorithms can be improved. CAS employs heuristic evaluation of the worth or merit of a subset of features. The merit function considers the usability of individual features for predicting the class label, along with the level of intercorrelation among them [6, 30] (equation 2).

CAS uses a correlation-based heuristic to evaluate the worth of attributes in a given model. This heuristic takes into account the usefulness of individual attributes for predicting the class label along with the level of intercorrelation among them [30]. Hall [30] indicated that the hypothesis on which the heuristic is based is "good feature subsets contain features highly correlated with the class, yet uncorrelated with each other." The workflow for the CAS attribute selection process is presented in Figure 2.

The algorithm was found to be fast to execute in a huge dataset (with 24 attributes and 519,000 records for each attribute) and was found to be helpful for managing the different attributes with 519,000 records. Compared to PCA, CAS was found to be much faster in our experimental analysis. CAS has a ranking approach using a GreedyStepwise search method [26, 30], which is suitable for selecting our drought attributes for the practical drought monitoring applications.
CAS filters correlation-based attribute selection scores by rewarding them for containing attributes that are highly correlated with the dependent variable and penalizing subsets for having attributes that are highly correlated among each other. A higher merit score represents a better subset [30].

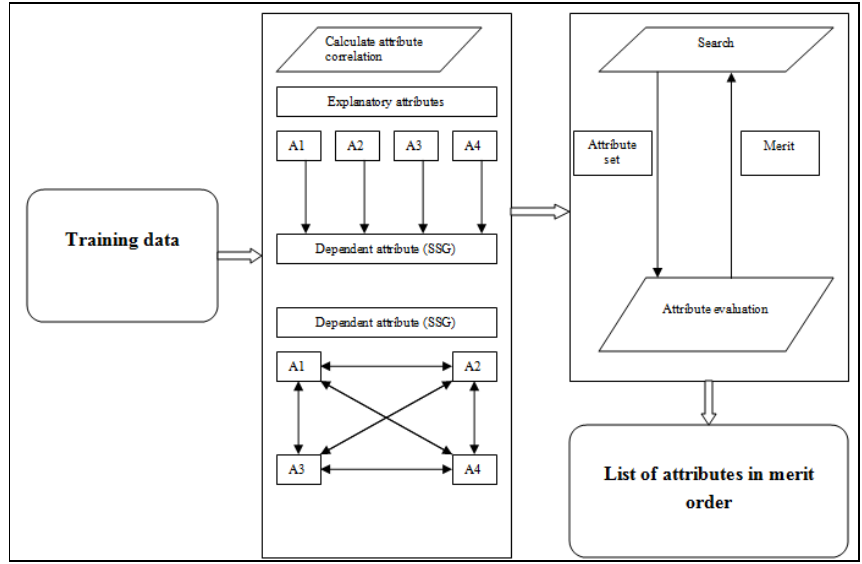

Figure 2. The process of a correlation-based attribute selection approach: $A_i$ represents attributes, and SSG is the standardized seasonal greenness dependent attribute (adapted from Hall [6]).

The CAS attribute selection algorithm uses a correlation-based measure to evaluate the worth of attribute subsets. This works with the principle that if a feature subset S contains k features, the evaluation measure for S is calculated as in equation 2 [2, 30].

$$Merit_s = \frac{k \overline{r_{cf}}}{\sqrt{k + k(k-1)\overline{r_{ff}}}} \quad (2)$$

where $Merit_s$ is the heuristic merit of attribute subsets $S$ containing $k$ attributes, $\overline{r_{cf}}$ is the average feature class correlation, and $\overline{r_{ff}}$ is the average feature-feature intercorrelation. Equation 2 is further described by Hall [30] for continuous data analysis. Equation 2 is, in fact, Pearson's correlation, where all variables have been standardized. The numerator can be thought of as giving an indication of how predictive a group of attributes are; the denominator can be considered an indicator of how much redundancy there is among them. The heuristic handles irrelevant features because they will be poor predictors of the class. Redundant attributes are discriminated against because they will be highly correlated with one or more of the other attributes. Hence, the merit function will have larger values for attribute subsets that have attributes with strong class–attribute correlation and weak attribute–attribute correlation. However, even if a set of attributes has strong class–attribute correlation, if there is strong attribute–attribute correlation, the merit value will be degraded [31].

The CAS algorithm treats continuous and discrete attributes differently. For continuous class data, the obvious measure for estimating the correlation between attributes is as in equation 2 [2, 30], which is a standard linear (Pearson's) correlation (equation 3) [30].

$$r_{xy} = \frac{\sum xy}{n\sigma_x \sigma_y} \quad (3)$$

International Journal of Data Mining & Knowledge Management Process (IJDKP) Vol.7, No.4, July 2017

where $r_{xy}$ is correlation values, n is number of attributes, $\sigma$ is standard deviation value, and $X$ and $Y$ are two continuous variables expressed in terms of deviations.

When one attribute is continuous and the other is discrete, a weighted Person's correlation is calculated, as shown in equation 4 [30]. Specifically for discrete attribute $X$ and a continuous attribute $Y$, if $X$ has $k$ values, then $k$ binary attributes are correlated with $Y$. Each of the $i = 1,...,k$ binary attributes takes value 1 when the $i^{th}$ value of $X$ occurs and 0 for all other values. Each of the $i = 1,...,k$ correlations calculated is weighted by the prior probability that $X$ takes value $i$.

$$r_{xy} = \sum_{i=1}^{k} p(X = x_i) r_{X_{bi}Y} \qquad (4)$$

where $r_{xy}$ is correlation values, $X_{bi}$ is a binary attribute that takes value 1 when $X$ has value $x_i$ and 0 otherwise.

Since the data types were defined as continuous (such as DEM, SPI, etc.) and discrete (such as landcover, ECO) (Figure 3), the CAS algorithm in Weka has properly produced the merit values as correlation values.

```
@relation GHA_dekad_22_out01

@attribute awc numeric
@attribute dem numeric
@attribute eco
{?,10085,10088,10089,10092,10093,10094,10096,10097,10108,10109,10
112,10113,10116,10118,10121,10122,10123,10125,10128,10129,10130,1
0131,10132,10133,10135,10136,10138,10142,10147,10151,10153,10159,
10161,10162,10167,10169,10174,10175,10176,10177,10179,10184,10189
,10190,10191,10194,10673,10695,10696,10698,10700,10702,17003}
@attribute landcover
{11,14,20,30,40,50,60,90,110,120,130,140,150,160,180,190,200,?}
@attribute amm numeric
@attribute amo numeric
@attribute best numeric
@attribute dmi numeric
@attribute mei numeric
@attribute nao numeric
@attribute Nino34 numeric
@attribute Nino4 numeric
@attribute oni numeric
@attribute pdo numeric
@attribute pna numeric
@attribute qbo numeric
@attribute sflux numeric
@attribute soi numeric
@attribute tin numeric
@attribute tna numeric
@attribute tsa numeric
@attribute Zscore numeric
@attribute SM numeric
@attribute SSG_dek22 numeric
@attribute SSG_dek23 numeric

@data
?,599,10698,200,410,190,510,21652,366,-220,2688,2905,-100,-770,-
140,-2167,1631,-700,-2846,340,140,-332,?,-1610,-1486
?,599,10698,200,-1990,119,1450,-
91815,949,360,2782,2948,900,600,610,1060,1839,-1600,-3001,-
50,10,-332,?,-991,-557
?,599,10698,200,1370,428,170,-11446,267,-220,2718,2908,200,880,-
310,-2464,1221,200,-2412,590,490,-332,?,867,1300
?,599,10698,200,3700,330,900,-39820,665,-
740,2770,2937,700,850,1520,874,1110,-500,-3253,680,220,-
332,?,247,371
```

Figure 3. An excerpt from the data type definition for CAS in java api. The ? mark is the missing data for the attributes' values.

In CAS, there are two major components for selecting the relevant attributes and ranking them according to their merits: attribute subset evaluation (CfsSubsetEval), and a search method. For the subset evaluation, we used the CfsSubsetEval algorithm, and for the search, the GreedyStepwise [26]. The GreedyStepwise search algorithm was found to be best for our purpose compared to the other two algorithms (BestFirst and ranker algorithms), since it has a ranking order for selecting the most relevant attributes. CfsSubsetEval algorithm [6, 26] evaluates the worth of a subset of attributes by considering the individual predictive ability of each feature along with the degree of redundancy between them; the GreedyStepwise [26] performs a greedy forward or backward search through the space of attribute subsets. The GreedyStepwise algorithm starts with no/all attributes or from an arbitrary point in the space, and it stops when the addition/deletion of any remaining attributes results in a decrease in evaluation. Finally, this algorithm produces a ranked list of attributes by traversing the space from one side to the other and recording the order that attributes are selected [5].

## 3. RESULTS AND DISCUSSIONS

The data used for the experiment have dekadal (10-day time interval) temporal resolutions. In a year, there are 36 dekads (since there are 3 dekads in a month). Representing these 36 dekads, 3 dekads were used for the attribute selection experiment: dekad 7, 22, and 31. Dekad 7 is March 1–March 10, dekad 22 is August 1–August 10, and dekad 31 is November 1–November 10. These dekads were selected systematically, representing the middle of the vegetation growth in the GHA. The middle of the growing season was selected with the assumption that there would be strong correlation between independent attributes and dependent attributes (SSG), which may help us in selecting the relevant attribute and discarding the irrelevant attributes.

In our empirical experiment for finding the relevant attribute for drought modeling, the PCA and CAS algorithms were found to be able to handle the continuous huge datasets with 519,000 records. The major outputs from PCA and CAS analysis are presented in the following subsections.

### 3.1 PCA ATTRIBUTE SELECTION

For the PCA analysis, the discrete attributes (eco and LC) were excluded and only the continuous data types were used. Detailed exploratory analysis was conducted for the intercorrelations within the identified attributes for modeling drought in the study area. The iterative analyses for PCA were done for the first dekads in May, August, and November. Since all of them showed same pattern, only the analysis output for the first dekad in August is presented here.

Table 2 presents the correlation matrix for the attributes used in predicting the second dekad in August using data from the first dekad in August. This correlation matrix shows how the attributes correlate with each other and which ones may show duplication of information if we use these attributes without attribute selection experimental analysis. For example, it was observed that "best" (Bivariate ENSO Timeseries) (https://www.esrl.noaa.gov/psd/data/climateindices/list/) has strong correlations with Nino34, Nino4, oni, and soi (Table 2). These intercorrelations range from 0.70 to 0.90. This duplication of information or redundant information directly led to the problem of overfitting and the overall model performance uncertainty [13]. In addition to the overfitting, Kumar and Batth [13] indicated that if there is too much redundant information present, then learning during the training phase is more difficult.

Although the choice of threshold for the variance is often selected heuristically and depends on the problem domain [28], in our experimental analysis, we set a constraint to capture 95% of the variance. Accordingly, 15 principal components (PCs) were needed to capture this threshold in our experimental attributes and datasets. Our intention here when using PCA as a tool for attribute selection is to select variables according to the magnitude of their coefficients (loadings). Table 3 presents the eigenvectors for the 15 PCs and for all attributes considered. From this table it is possible to observe the eigenvectors (the coefficient values) for each attribute and the contributions of the attributes in each identified PC (a total 15 PCs here).

Table 4 presents the eigenvalues for the selected PCs for capturing the 95% variability in decreasing order of the eigenvalues, proportions, and cumulative variability. The eigenvalues here show the variance in all the variables (rows), which is accounted for by the selected features. In principle, the eigenvalues measure the amount of variation in the total sample accounted for by each feature selected in explaining the class attribute.

The proportion in the second column of Table 4 shows the ratio of eigenvalues, which is the ratio of explanatory importance of the selected features with respect to the other features. If a factor has a low eigenvalue, then it is contributing little to the explanation of variances in the variables and may be ignored as redundant with more important factors. The first selected features explained about 30%, the second about 13%, and the third about 7% of the overall variability. To achieve 95% of the variance, 14 PCs were needed (Table 4). The challenge here is that in each PC, all the attributes were included even though they have very low eigenvector (coefficient) values.

Tables 2-4 clearly showed which attributes are related to the class attribute (SSG) and also the duplications of information available within the selected attributes. The PCA analysis here helped us to explore the relationships between the selected attributes to explain the class attributes and also the duplication of information if we use all the attributes identified for modelling the vegetation condition (SSG) target values. In the following subsections, the CAS attribute selection with GreedyStepwise algorithm prioritizes each attribute based on their strong correlations between class attribute SSG and also their low information duplication within the explanatory attributes.

Table 2: Correlation matrix for first August dekad outlook 1. Acronyms listed here are spelled out in Table 1. The same pattern was observed for May and November (not presented here).

| | awc | dem | amm | amo | best | dmi | mei | nao | Nino34 | Nino4 | oni | pdo | pna | qbo | sflux | soi | tin | tna | tsa | N_Precip | SM | SSG |
|---|---|---|---|---|---|---|---|---|---|---|---|---|---|---|---|---|---|---|---|---|---|---|
| awc | 1 | | | | | | | | | | | | | | | | | | | | | |
| dem | 0.16 | 1 | | | | | | | | | | | | | | | | | | | | |
| amm | 0 | 0 | 1 | | | | | | | | | | | | | | | | | | | |
| amo | 0 | 0 | 0.77 | 1 | | | | | | | | | | | | | | | | | | |
| best | 0 | 0 | 0.2 | 0.4 | 1 | | | | | | | | | | | | | | | | | |
| dmi | 0 | 0 | -0.35 | -0.48 | -0.57 | 1 | | | | | | | | | | | | | | | | |
| mei | 0 | 0 | -0.01 | 0.13 | 0.82 | -0.35 | 1 | | | | | | | | | | | | | | | |
| nao | 0 | 0 | -0.2 | -0.2 | -0.06 | -0.01 | 0.17 | 1 | | | | | | | | | | | | | | |
| Nino34 | 0 | 0 | 0.31 | 0.41 | 0.93 | -0.54 | 0.81 | -0.11 | 1 | | | | | | | | | | | | | |
| Nino4 | 0 | 0 | 0.21 | 0.35 | 0.82 | -0.38 | 0.7 | -0.01 | 0.78 | 1 | | | | | | | | | | | | |
| oni | 0 | 0 | 0.24 | 0.33 | 0.9 | -0.52 | 0.87 | -0.01 | 0.98 | 0.74 | 1 | | | | | | | | | | | |
| pdo | 0 | 0 | -0.12 | -0.12 | 0.59 | -0.31 | 0.44 | 0.2 | 0.53 | 0.68 | 0.53 | 1 | | | | | | | | | | |
| pna | 0 | 0 | 0.22 | 0.44 | 0.64 | -0.4 | 0.54 | 0.29 | 0.49 | 0.45 | 0.45 | 0.36 | 1 | | | | | | | | | |
| qbo | 0 | 0 | -0.17 | 0.07 | -0.08 | -0.04 | -0.24 | 0.03 | -0.24 | -0.26 | -0.17 | -0.02 | 0.14 | 1 | | | | | | | | |
| sflux | 0 | 0 | -0.37 | 0.01 | 0.26 | -0.35 | 0.4 | 0.25 | 0.29 | 0.18 | 0.33 | 0.2 | 0.35 | 0.12 | 1 | | | | | | | |
| soi | 0 | 0 | 0.31 | 0.09 | -0.66 | 0.14 | -0.86 | -0.06 | -0.66 | -0.54 | -0.7 | -0.42 | -0.3 | 0.34 | -0.35 | 1 | | | | | | |
| tin | 0 | 0 | -0.1 | -0.28 | -0.85 | 0.47 | -0.76 | 0.03 | -0.87 | -0.74 | -0.88 | -0.55 | -0.33 | 0.14 | -0.35 | 0.68 | 1 | | | | | |
| tna | 0 | 0 | 0.54 | 0.76 | 0.47 | -0.59 | 0.16 | -0.42 | 0.48 | 0.26 | 0.43 | -0.1 | 0.23 | 0.28 | -0.03 | -0.01 | -0.34 | 1 | | | | |
| tsa | 0 | 0 | 0.32 | 0.17 | 0.19 | -0.13 | 0.03 | -0.65 | 0.25 | 0.01 | 0.17 | -0.13 | 0.04 | -0.08 | -0.17 | 0.01 | 0.04 | 0.4 | 1 | | | |
| N_Precip | 0.01 | 0.03 | 0.12 | 0.05 | -0.03 | -0.07 | -0.08 | 0 | 0 | 0.01 | -0.04 | 0.01 | -0.03 | -0.15 | -0.1 | 0.07 | 0.02 | 0.02 | -0.09 | 1 | | |
| SM | 0.07 | 0.17 | 0.01 | 0 | 0 | -0.02 | 0 | 0 | 0 | 0.01 | 0 | 0 | 0 | -0.03 | -0.01 | 0 | 0 | 0.01 | -0.01 | 0.14 | 1 | |
| SSG | 0 | 0 | 0.08 | 0.17 | 0.18 | -0.12 | 0.06 | -0.01 | 0.14 | 0.14 | 0.11 | 0.08 | 0.12 | -0.02 | 0.01 | -0.07 | -0.17 | 0.14 | -0.02 | 0.09 | 0.03 | 1 |

Table 3: Eigenvectors for the analyzed attributes vs the 15 PCs (for first August dekad outlook 1).

|      | PC 1    | PC 2    | PC 3    | PC 4    | PC 5    | PC 6    | PC 7    | PC 8    | PC 9    | PC 10   | PC 11   | PC 12   | PC 13   | PC 14   | PC 15   |
|------|---------|---------|---------|---------|---------|---------|---------|---------|---------|---------|---------|---------|---------|---------|---------|
| awc  | 0       | -0.0008 | 0.0014  | 0.201   | -0.0734 | -0.7463 | 0.0922  | -0.2216 | -0.2941 | -0.3173 | -0.337  | 0.0999  | 0.0783  | 0.1508  | -0.0016 |
| dem  | -0.0001 | -0.0018 | 0.0039  | 0.4261  | -0.1503 | -0.4186 | -0.0186 | -0.0373 | 0.1506  | 0.2995  | 0.6449  | -0.1784 | -0.1068 | -0.2176 | -0.0011 |
| amm  | -0.0947 | -0.4401 | 0.0272  | 0.0506  | 0.3293  | -0.045  | -0.3011 | -0.0848 | 0.0107  | 0.0522  | -0.018  | 0.0433  | 0.0756  | -0.0482 | -0.0594 |
| amo  | -0.165  | -0.4061 | 0.2361  | 0.0103  | 0.1231  | -0.0176 | -0.1305 | 0.0019  | -0.1398 | 0.1276  | 0.0004  | 0.1239  | -0.1287 | 0.0246  | 0.3697  |
| best | -0.3593 | -0.0095 | -0.0123 | -0.0091 | -0.0386 | 0.0049  | 0.032   | -0.0721 | 0.1009  | -0.0314 | -0.0382 | -0.0815 | -0.0895 | 0.0358  | -0.0384 |
| dmi  | 0.229   | 0.1674  | -0.2635 | -0.0045 | 0.0218  | 0.0012  | -0.0666 | -0.2274 | 0.0341  | 0.1607  | -0.1342 | -0.0595 | -0.5269 | 0.1589  | 0.3571  |
| mei  | -0.3187 | 0.1801  | -0.0806 | -0.0113 | -0.0406 | -0.0159 | -0.1312 | 0.0675  | -0.145  | 0.1115  | -0.0621 | -0.07   | -0.2084 | 0.0628  | -0.1752 |
| nao  | 0.0037  | 0.3257  | 0.4289  | 0.0505  | 0.2908  | -0.0375 | -0.2611 | -0.0326 | -0.0772 | 0.1011  | -0.0361 | -0.0048 | 0.0018  | -0.0188 | -0.4315 |
| Nino34 | -0.3576 | -0.0338 | -0.1144 | 0      | 0.0026  | -0.0045 | -0.0169 | 0.0252  | -0.0083 | -0.0125 | 0.0159  | 0.025   | 0.0053  | 0.002   | -0.0631 |
| Nino4 | -0.3114 | 0.0473 | -0.0746 | 0.0304  | 0.1657  | -0.0061 | -0.0326 | -0.1795 | 0.2039  | -0.0671 | 0.0029  | 0.0696  | 0.033   | -0.0297 | 0.4265  |
| oni  | -0.3528 | 0.0212  | -0.0846 | -0.0103 | -0.0359 | -0.0056 | -0.0236 | 0.0383  | -0.0218 | -0.0081 | 0.0291  | 0.1031  | -0.0588 | 0.0094  | -0.198  |
| pdo  | -0.2165 | 0.2276  | 0.0314  | 0.0219  | 0.0961  | 0.0074  | 0.1073  | -0.2632 | 0.5098  | -0.2964 | 0.015   | -0.0921 | 0.2775  | -0.0751 | 0.0582  |
| pna  | -0.2259 | -0.0017 | 0.3594  | -0.0094 | 0.0184  | -0.0272 | -0.2185 | -0.0773 | 0.0134  | 0.1008  | -0.2216 | -0.5797 | -0.1238 | 0.0924  | 0.1385  |
| qbo  | 0.0602  | -0.0762 | 0.4719  | -0.0992 | -0.425  | 0.0307  | 0.2331  | -0.1212 | 0.3051  | -0.1693 | -0.024  | 0.0232  | -0.3665 | 0.1137  | -0.1067 |
| sflux | -0.1359 | 0.2267 | 0.2958  | -0.0553 | -0.298  | 0.0087  | 0.0895  | 0.3681  | -0.322  | 0.0276  | 0.071   | -0.0738 | 0.3201  | -0.0644 | 0.4329  |
| soi  | 0.2562  | -0.2706 | 0.2444  | 0.0128  | 0.1025  | -0.0103 | -0.0375 | -0.0895 | 0.1856  | -0.1175 | 0.0281  | 0.0129  | 0.2479  | -0.0872 | 0.0941  |
| tin  | 0.328   | -0.0764 | 0.0593  | -0.0017 | 0.0104  | -0.0162 | -0.126  | -0.0007 | -0.0067 | 0.0331  | -0.1057 | -0.3419 | 0.073   | 0.0088  | -0.0314 |
| tna  | -0.1751 | -0.4223 | 0.1153  | -0.0381 | -0.1805 | 0.016   | 0.1321  | 0.0884  | -0.0302 | -0.0403 | 0.0642  | 0.1917  | -0.1731 | 0.0461  | -0.1566 |
| tsa  | -0.0541 | -0.3133 | -0.3655 | -0.0703 | -0.3102 | 0.002   | -0.0123 | 0.0333  | 0.0372  | -0.0188 | -0.1352 | -0.521  | 0.2034  | -0.0157 | -0.1656 |
| N_Precip | 0.0063 | -0.0532 | -0.007 | 0.2075 | 0.4749 | 0.0032 | 0.3667 | 0.4479 | -0.0123 | -0.3624 | 0.1703 | -0.2933 | -0.2088 | 0.3027 | 0.0107 |
| SM   | -0.0023 | -0.0089 | 0.0047  | 0.5435  | -0.0398 | 0.1227  | 0.0639  | 0.2801  | 0.1426  | 0.0085  | -0.5013 | 0.0671  | -0.1556 | -0.5518 | 0.0134  |
| SSG_dek22 | -0.0676 | -0.0587 | 0.0837 | 0.083 | 0.1881 | 0.1492 | 0.6735 | -0.4098 | -0.241 | 0.4317 | -0.0902 | -0.1116 | 0.1272 | -0.0512 | -0.0726 |

Table 4: Eigenvalue for the 14 PCs (for first August dekad outlook 1).

| Principal Components (PC) | Eigenvalue | Proportion | Cumulative Variance |
|---------------------------|------------|------------|---------------------|
| PC 1                      | 7.26538    | 0.30272    | 0.30272             |
| PC 2                      | 3.16231    | 0.13176    | 0.43449             |
| PC 3                      | 1.75295    | 0.07304    | 0.50753             |
| PC 4                      | 1.64945    | 0.06873    | 0.57625             |
| PC 5                      | 1.3762     | 0.05734    | 0.6336              |
| PC 6                      | 1.06352    | 0.04431    | 0.67791             |
| PC 7                      | 1.00475    | 0.04186    | 0.71977             |
| PC 8                      | 0.93827    | 0.03909    | 0.75887             |
| PC 9                      | 0.89764    | 0.0374     | 0.79627             |
| PC 10                     | 0.87402    | 0.03642    | 0.83269             |
| PC 11                     | 0.77798    | 0.03242    | 0.8651              |
| PC 12                     | 0.72802    | 0.03033    | 0.89544             |
| PC 13                     | 0.64466    | 0.02686    | 0.9223              |
| PC 14                     | 0.60315    | 0.02513    | 0.94743             |

### 3.2 CAS ATTRIBUTE SELECTION

For this analysis, outlook prediction data for the first dekads in May, August, and November were used. Data from these three dekadal outlook predictions were used because they are in the middle of the growing season in the study area and also are assumed to have higher prediction accuracies than other dekadal (ten-days interval) time-lag prediction.

In this experimental analysis, the attribute evaluator was CfsSubsetEval and the corresponding search method was the GreedyStepwise algorithm. The attribute evaluator is the technique by which each attribute in our dataset is evaluated in the context of the dependent variable (SSG). The search method is the technique by which to navigate different combinations of attributes in the dataset in order to arrive on a short list of chosen attributes. The CfsSubsetEval algorithm evaluates the worth of a subset of attributes by considering the individual predictive ability of each attribute along with the degree of redundancy between the selected attributes [30]. The attributes that are highly correlated with the class attribute SSG while having low intercorrelation within themselves are the ones that we search for and subsequently rank by their merit order.

Table 5 presents dekad 7, dekad 22, and dekad 31 attribute lists with their merit rank for outlooks 1-5, respectively. The merit value here is in terms of correlation values, based on the hypothesis that good attribute subsets contain attributes highly correlated with the dependent attribute SSG, yet uncorrelated with each other [30].

Table 5: List of attributes for modeling drought in GHA. The merit value for the attributes is the correlation value (attribute-to-class SSG and attribute-to-attribute as indicated in equations 2, 3, and 4).

| Attribute name | March merit value | | | | | August merit value | | | | | November merit value | | | | |
|---|---|---|---|---|---|---|---|---|---|---|---|---|---|---|---|
| | Out1 | Out2 | Out3 | Out4 | Out5 | Out1 | Out2 | Out3 | Out4 | Out5 | Out1 | Out2 | Out3 | Out4 | Out5 |
| amm | 0.2593 | 0.2642 | 0.2674 | 0.2684 | 0.269 | 0.282 | 0.288 | 0.291 | 0.304 | 0.291 | 0.2968 | 0.3053 | 0.3013 | 0.2968 | 0.2931 |
| amo | 0.9717 | 0.9598 | 0.943 | 0.923 | 0.9072 | 0.946 | 0.938 | 0.938 | 0.924 | 0.911 | 0.9765 | 0.9735 | 0.9684 | 0.9616 | 0.9536 |
| awc | 0.9567 | 0.9449 | 0.9159 | 0.8973 | 0.8771 | 0.96 | 0.95 | 0.925 | 0.912 | 0.898 | 0.9635 | 0.9605 | 0.9556 | 0.9488 | 0.9409 |
| best | 0.2915 | 0.296 | 0.2986 | 0.3211 | 0.3198 | 0.385 | 0.391 | 0.393 | 0.391 | 0.388 | 0.4079 | 0.4008 | 0.393 | 0.3585 | 0.3553 |
| dem | 0.6781 | 0.6696 | 0.6627 | 0.6523 | 0.6382 | 0.692 | 0.688 | 0.682 | 0.674 | 0.667 | 0.7312 | 0.7281 | 0.7239 | 0.7182 | 0.7116 |
| dmi | 0.0846 | 0.0839 | 0.083 | 0.0829 | 0.0854 | 0.172 | 0.177 | 0.182 | 0.185 | 0.189 | 0.0935 | 0.0892 | 0.0851 | 0.0813 | 0.0775 |
| eco | 0.5659 | 0.5653 | 0.5639 | 0.5577 | 0.5468 | 0.615 | 0.612 | 0.606 | 0.599 | 0.593 | 0.6012 | 0.6474 | 0.6437 | 0.6387 | 0.6329 |
| landcover | 0.9874 | 0.9677 | 0.9501 | 0.9291 | 0.9013 | 0.989 | 0.978 | 0.964 | 0.948 | 0.933 | 0.9953 | 0.9915 | 0.9856 | 0.9675 | 0.9592 |
| mei | 0.348 | 0.3503 | 0.3505 | 0.3482 | 0.3452 | 0.426 | 0.431 | 0.432 | 0.43 | 0.426 | 0.5523 | 0.5475 | 0.5413 | 0.5349 | 0.5273 |
| nao | 0.4615 | 0.4618 | 0.4592 | 0.4529 | 0.4449 | 0.36 | 0.365 | 0.366 | 0.365 | 0.362 | 0.4488 | 0.47 | 0.4659 | 0.4613 | 0.4555 |
| Nino34 | 0.9415 | 0.9311 | 0.9295 | 0.9107 | 0.8902 | 0.993 | 0.982 | 0.967 | 0.952 | 0.937 | 0.949 | 0.9457 | 0.9406 | 0.9338 | 0.9258 |
| Nino4 | 0.9911 | 0.978 | 0.9601 | 0.9388 | 0.9156 | 0.975 | 0.971 | 0.958 | 0.943 | 0.928 | 0.9913 | 0.9873 | 0.9813 | 0.9735 | 0.9651 |
| oni | 0.5054 | 0.5046 | 0.5019 | 0.4962 | 0.5076 | 0.563 | 0.561 | 0.557 | 0.552 | 0.547 | 0.5198 | 0.5144 | 0.5079 | 0.5012 | 0.4933 |
| pdo | 0.3177 | 0.3209 | 0.322 | 0.2987 | 0.2984 | 0.316 | 0.319 | 0.32 | 0.319 | 0.301 | 0.3322 | 0.3266 | 0.3202 | 0.3141 | 0.3099 |
| pna | 0.4073 | 0.4065 | 0.4023 | 0.4191 | 0.4153 | 0.473 | 0.478 | 0.478 | 0.477 | 0.473 | 0.6504 | 0.5965 | 0.5906 | 0.5838 | 0.5761 |
| qbo | 0.4342 | 0.4324 | 0.4267 | 0.3956 | 0.388 | 0.299 | 0.304 | 0.305 | 0.292 | 0.315 | 0.3525 | 0.3697 | 0.3639 | 0.3876 | 0.3851 |
| sflux | 0.7437 | 0.7347 | 0.7272 | 0.7159 | 0.7005 | 0.754 | 0.75 | 0.743 | 0.735 | 0.727 | 0.7951 | 0.7918 | 0.7872 | 0.781 | 0.7738 |
| SM | 0.9804 | 0.9743 | 0.9567 | 0.9355 | 0.9124 | 0.982 | 0.965 | 0.951 | 0.936 | 0.922 | 0.9837 | 0.9803 | 0.9747 | 0.978 | 0.9699 |
| soi | 0.3797 | 0.3802 | 0.3781 | 0.3736 | 0.3684 | 0.26 | 0.265 | 0.267 | 0.267 | 0.265 | 0.3744 | 0.347 | 0.3407 | 0.3345 | 0.3303 |
| SSG_dek22 | 0.9944 | 0.9802 | 0.9611 | 0.9387 | 0.9145 | 0.995 | 0.984 | 0.968 | 0.952 | 0.937 | 0.9986 | 0.9948 | 0.9888 | 0.9811 | 0.9726 |
| tin | 0.5335 | 0.5306 | 0.5253 | 0.5171 | 0.4883 | 0.336 | 0.339 | 0.34 | 0.339 | 0.336 | 0.3097 | 0.2938 | 0.2903 | 0.2864 | 0.283 |
| tna | 0.9038 | 0.8947 | 0.881 | 0.8641 | 0.8453 | 0.828 | 0.821 | 0.811 | 0.8 | 0.79 | 0.8605 | 0.8572 | 0.8529 | 0.847 | 0.8402 |
| tsa | 0.8257 | 0.8205 | 0.8107 | 0.7969 | 0.7805 | 0.926 | 0.917 | 0.904 | 0.891 | 0.878 | 0.9349 | 0.9311 | 0.9256 | 0.9189 | 0.9113 |
| N_Precip | 0.6053 | 0.605 | 0.6039 | 0.5976 | 0.586 | 0.516 | 0.517 | 0.516 | 0.512 | 0.508 | 0.4787 | 0.4418 | 0.4342 | 0.4267 | 0.4238 |

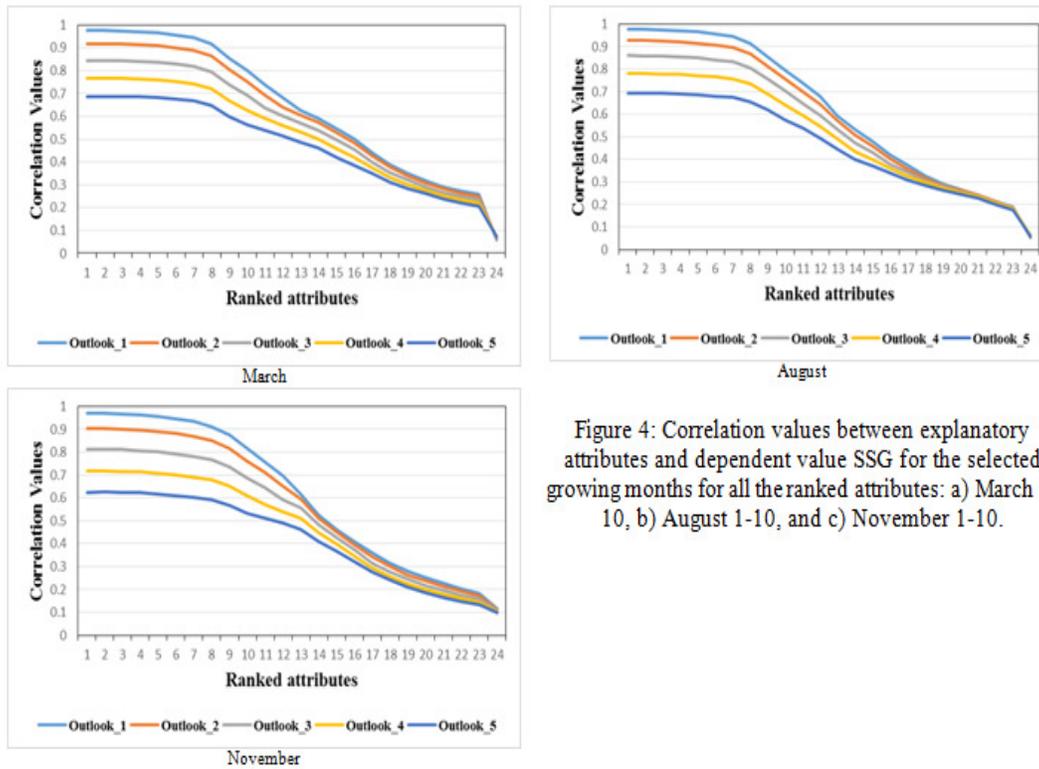

Figure 4: Correlation values between explanatory attributes and dependent value SSG for the selected growing months for all the ranked attributes: a) March 1-10, b) August 1-10, and c) November 1-10.

Table 5 for March (dekad 7 or March 1–10) presents the merits of the attributes (the correlation with dependent attribute SSG). The higher the merit, the more relevant the selected attribute. In Table 5, the merit value ranges from 0.08 to 0.99 for outlook 1, from 0.08 to 0.98 for outlook 2, from 0.08 to 0.96 for outlook 3, from 0.08 to 0.94 for outlook 4, and from 0.08 to 0.92 for outlook 5. As the outlook periods increase, the merit value was found to be decreasing. This is in line with our expectation in that as the prediction length increases, the vegetation condition (SSG) to be predicted is different from the predictors. For the first March dekad (March 1–10), outlook 1 is March 11–20 vegetation conditions (SSG), whereas outlook 5 is dekad 12 (April 20–30) vegetation conditions. In Table 5, a total of 15 attributes were found to have >0.5 merit value, and the remaining 10 attributes were found to have <0.5 for outlooks 1, 2, and 3. For outlooks 4 and 5, a total of 14 attributes were found to have >0.5 merit value. The correlations were found to decrease as the outlook period increased in the time lag predictions. Consistently, the best attributes were found to be the same for all the outlooks assessed in the time lag predictions.

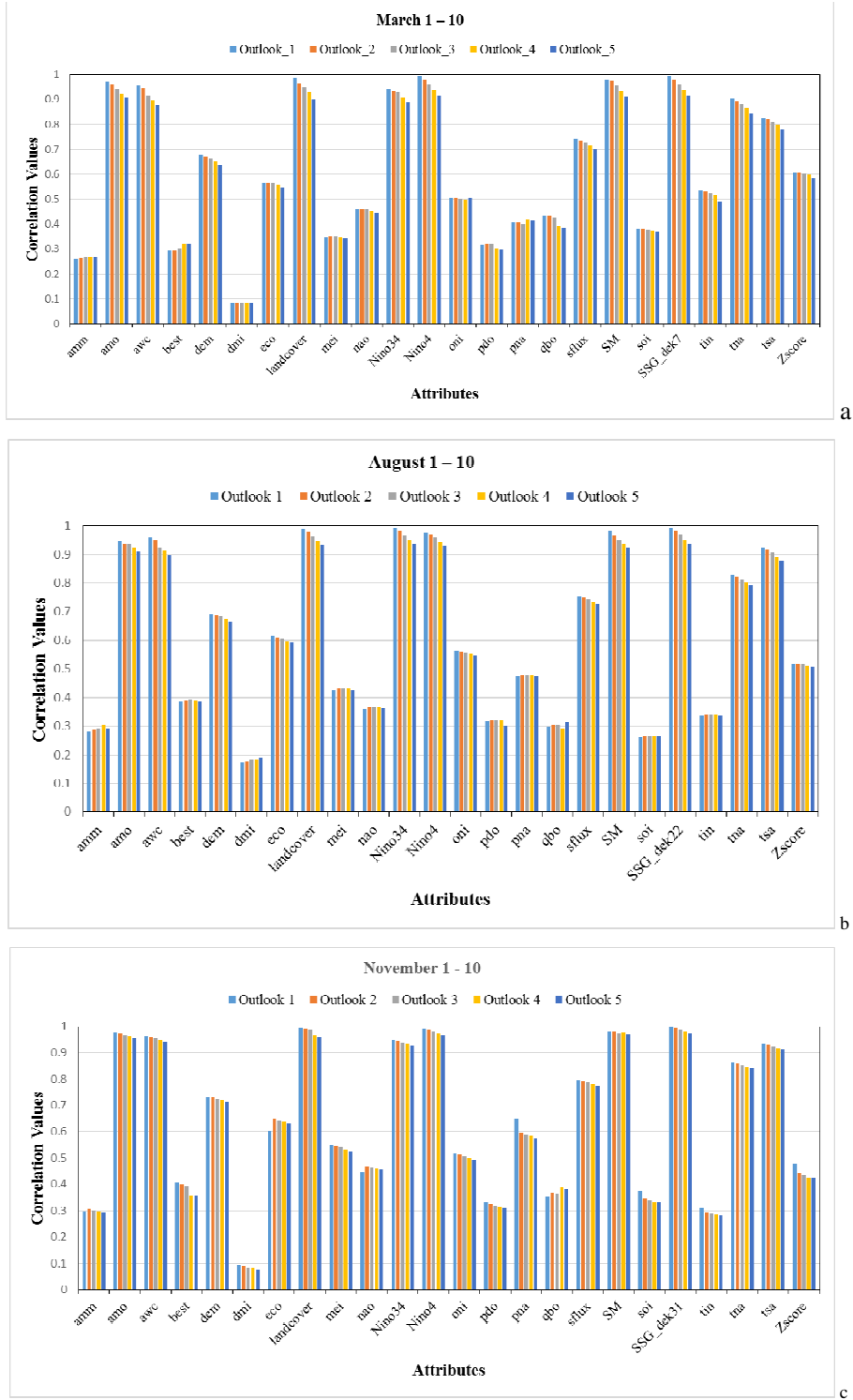

Figure 5: Correlation values between explanatory attributes and dependent value SSG for the selected growing months: a) March 1-10, b) August 1-10, and c) November 1-10.

With the same procedure used for the first dekad in May, the experimental analysis was repeated for the first dekads in August and November (Table 5). The same patterns were observed for all three dekadal periods, confirming the applicability of the CAS attribute selection approach for our domain area.

Figure 4 presents the correlation patterns of the attributes for March, August, and November under outlooks 1-5. As expected for the three growing periods (March, August, and November), outlook 1 had the highest correlation values and outlook 5 had the lowest correlation values. Figure 5a-c presents the correlation of each attribute with the target attribute SSG for the March, August, and November growing periods. In all of the assessed attributes, as the prediction time-lag increases, the correlations were found to decrease.

## 4. CONCLUSIONS

In this research, we developed an attribute selection approach with special emphasis for drought modeling and prediction. The empirical study presented here confirmed that the automated data mining attribute selection technique is an objective-based attribute selection approach compared to the subjective attribute selection approaches, which rely on past study reviews, common sense, and theory-based consultation.

From the experimental analysis using real world drought data, we developed an empirical method for selecting relevant attributes for modeling drought and also selected the most relevant attribute for drought modeling and predictions in the GHA. The list of attributes with a relevancy threshold value of merit >0.5 for the three growing months (May, August, and November) are presented in Table 6. The experimental outputs were also evaluated through experts' assessment of the domain-specific descriptions of the nature of the attributes for their relevance to vegetation conditions and drought modeling in the study area. This experimental evaluation is ongoing.

Table 6: List of relevant attributes with their merit values for modeling drought in GHA. The average merit value is for the middle of the three growing seasons (March, August, and November), outlook 1–5, as presented in Table 5.

| Attribute name | Abbreviation | Average Merit value | Rank |
|---|---|---|---|
| Atlantic Multi-decadal Oscillation | amo | 0.946353 | 5 |
| Available Water Holding Capacity of the Soil | awc | 0.933747 | 7 |
| Digital Elevation Model | dem | 0.687793 | 11 |
| Ecoregion | eco | 0.599233 | 12 |
| Land Cover | landcover | 0.963113 | 3 |
| East Central Tropical Pacific SST | Nino34 | 0.941927 | 6 |
| Central Tropical Pacific SST | Nino4 | 0.963807 | 2 |
| Normalized Precipitation | N_Precip | 0.518133 | 14 |
| Oceanic Nino Index | oni | 0.522153 | 13 |
| Pacific North American Index | pna | 0.495127 | 15 |
| Solar Flux (10.7cm) | sflux | 0.75066 | 10 |
| Soil Moisture | SM | 0.960127 | 4 |
| Standardized Seasonal Greenness | SSG | 0.97072 | 1 |
| Tropical Southern Atlantic Index | tna | 0.846447 | 9 |
| Tropical Southern Atlantic Index | tsa | 0.87814 | 8 |

In conclusion, in the experimental analysis that we did on four attribute selection approaches (CAS, PCA, ReliefAttributEval, and WraperSubsetEval), the CAS and PCA were found to be helpful. The other two options (ReliefAttributEval and WraperSubsetEval) were found to be computationally intensive and impractical with normal machines, since they need high computing machines with multicore processor capacities.

Future research may use the ReliefAttributEval and WraperSubsetEval algorithms for improved results on attribute search processes. For the current experiment, it was found that these two algorithms were computationally intensive, and we could not execute the experimental analysis. Multicore processing machines can be used for this challenge to get the experimental outputs in reasonable time. Future research may also evaluate the developed methodology using relevant classification techniques and quantify the actual information gain from the developed approach.

## ACKNOWLEDGMENTS

This work was supported by NASA Project NNX14AD30G. The authors are grateful to Deborah Wood of the National Drought Mitigation Center for her editorial comments.